\documentclass[aps,prd,eqsecnum,showpacs,superscriptaddress,twocolumn,
nofootinbib, floatfix,preprintnumbers,altaffilletter]{revtex4}

\usepackage{graphicx,hyperref,amssymb,amsmath,longtable,rotating,color,epsfig,
epsf,fancyhdr}

\newcommand{\ee}[1]{\!\times\!10^{#1}}
\newcommand{\gws}{gravitational waves~}
\newcommand{\gw}{gravitational wave~}

\newcommand{\tempo}{T\textsc{empo}~}

\begin{document}

\pagestyle{fancy}

\title{Binary system delays and timing noise in searches for gravitational waves
from known pulsars}
\author{Matthew Pitkin and Graham Woan}
\affiliation{Department of Physics and Astronomy, University of Glasgow, University Avenue,
Glasgow, G12 8QQ, UK}
\email{matthew@astro.gla.ac.uk}

\date{\today}

\begin{abstract}
The majority of fast millisecond pulsars are in binary systems, so
that any periodic signal they emit is modulated by both Doppler and
relativistic effects. Here we show how well-established binary
models can be used to account for these effects in searches for
gravitational waves from known pulsars within binary systems. A seperate 
issue affecting certain pulsar signals is that of timing noise and we show how 
this, with particular reference to the Crab pulsar, can be compensated 
for by using regularly updated timing ephemerides.
\end{abstract}

\preprint{LIGO-P060063-02-Z}

\maketitle

\section{Introduction}\label{intro}
Neutron stars are thought to be strong candidates for the emission
of detectable continuous \gws \cite{Schutz:1989}, including the 1627
pulsars currently discovered\footnote{As given by the Australia
Telescope National Facility - ATNF - online pulsar catalogue
\cite{ATNF, Manchester:2005} as of $6^{\rm th}$ Nov 2006, from which
time all subsequent pulsar numbers will be taken.}. The majority of
these pulsars have been discovered through dedicated radio surveys.
Surveys are ongoing, but estimates of the number of active pulsars in the galaxy
can be made by inference from the current population, taking into
account biasing from selection effects, and the supernova rate.
Estimates give values of $\sim 200\,000$ active pulsars within our
galaxy (see Ref.~\cite{Lorimer:2001}).

Pulsars are found in a wide range of environments. Some are directly
associated with the supernova remnants (SNRs) in which they were
born. These are typically young pulsars whose birth velocity has not
yet caused a large displacement from the remnant and the SNR has not
dissipated into the interstellar medium (ISM). Some pulsars are
found in binary systems as the companions of a wide range of
astronomical bodies from planets, through main sequence stars, to
white dwarfs and other neutron stars. The fastest, `millisecond',
pulsars (pulsars with rotation periods of $< 10$\,ms) are usually
found within binary systems, and often within globular clusters,
their rapid rotation rate a consequence of being spun-up by
accretion of material from a stellar companion \cite{vandenHeuvel:1984}.  
Pulsars are also seen without any association, and in this paper 
we will classify any pulsar not in a binary system as \emph{isolated}.

Pulsars are generally seen to spin-down as they lose rotational
energy through a variety of emission mechanisms, but the primary loss
mechanism is thought to be magnetic dipole radiation. Other
potential mechanisms include particle acceleration and gravitational
radiation. Whatever the mechanisms at work, the rotational phase
evolution of a pulsar can generally be well described by a short
Taylor expansion,
\begin{eqnarray}\label{PhaseTaylorExp}
\phi(T) & = & \phi_0 + 2\pi{}\bigg\{\nu_0(T-t_0) + \frac{1}{2}\dot{\nu_0}(T-t_0)^2 \nonumber \\
 & & + \frac{1}{6}\ddot{\nu_0}(T-t_0)^3 + \ldots\bigg\},
\end{eqnarray}
where $\phi_0$ is the initial phase, $\nu_0$ and its time
derivatives are the pulsar frequency and spin-down coefficients at
an epoch $t_0$, and $T$ is the time in a frame comoving with the
pulsar. For the vast majority of pulsars the value of $\dot{\nu}$ is
very small and $\ddot{\nu}$ is unmeasurable or swamped by timing
noise (see \S\ref{TimingNoise}). Note that \gws emitted from a
triaxial, non-precessing, neutron star come from the quadrupolar
component of the rotating body and so will have exactly \emph{twice}
the phase evolution described by Eq.~\ref{PhaseTaylorExp}.

The expected gravitational wave signal from a triaxial neutron star
is given by \cite{JKS:1998}
\begin{eqnarray}\label{PulsarSignal}
h(t) & = & \frac{1}{2}F_+(t;\psi)h_0(1+\cos^2\iota)\cos{2\phi(t)} \nonumber \\
 & & + F_{\times}(t;\psi)h_0\cos{\iota}\sin{2\phi(t)},
\end{eqnarray}
where $\phi(t)$ is that given in Eq.~(\ref{PhaseTaylorExp}), $F_+$
and $F_{\times}$ are the detector beam patterns for the plus and
cross polarisations of the gravitational waves, $\psi$ is the wave
polarisation angle, and $\iota$ is the angle between the rotation
axis of the pulsar and the line-of-sight. For a gravitational wave
signal impinging on the Earth the signal arrival time at the
detector, $t$, will be modulated by Doppler, time delay and
relativistic effects caused by the motions of the Earth and other
bodies in the solar system, so we want a reference frame which will
not be affected by these. In general, for isolated
pulsars\footnote{assuming negligible proper motion and acceleration
within a globular clusters.}, such a frame is the solar system
barycentre (SSB). To convert from $t$ to the time at the SSB, $t_b$,
we must include a series of time corrections,
\begin{equation}\label{TimeDelay}
t_b = t + \delta{}t = t + \frac{\mathbf{r}\cdot\hat{\mathbf{n}}}{c} + \Delta_{E_{\odot}} +
\Delta_{S_{\odot}},
\end{equation}
where $\mathbf{r}$ is the position of the detector with respect to
the solar system barycentre (SSB), $\hat{\mathbf{n}}$ is the unit
vector pointing to the pulsar, $\Delta_{E_{\odot}}$ is the special
relativistic Einstein delay, and $\Delta_{S_{\odot}}$ is the general
relativistic Shapiro delay (see Ref.~\cite{TaylorWeisberg:1989} for
definitions of these delay terms). For isolated pulsars we can
therefore set $T=t_b$ in Eq.~(\ref{PhaseTaylorExp}). For pulsars in
binary systems there will be additional time delays as discussed in
\S\ref{Binaries}.

In this paper we describe the addition of such binary system time
delays with respect to current searches for \gws from known pulsars.
In a seperate issue we also discuss how pulsar timing noise can 
cause deviations from the simple phase model in Eq.~\ref{PhaseTaylorExp}, 
and how to account for this in the analysis.

\section{Search method}\label{sec:method}
Searches for \gws from a selection of known pulsars have been
performed using data from the LIGO \cite{Abbott:2003} and GEO\,600
\cite{Willke:2002} \gw detectors from the four science runs (S1-4),
which have taken place since late 2002 \cite{Abbott:2004,
Abbott:2005, Abbott:2006}. The search method used is outlined
briefly here, but is described more fully in
Refs.~\cite{Dupuis:2004, DupuisWoan:2005}.

All current data from these detectors is sampled at 16\,384\,Hz,
giving a range of 0--8192\,Hz available for searches, although in 
practice this is limited to the most sensitive region of the 
detectors between $\approx 50$--4000\,Hz. The rotation
frequencies of known pulsars are, of course, known very precisely
from radio and/or X-ray observations, so the vast majority of this
frequency space is redundant and the speed of any search can be
increased by removing it. Knowledge of the pulsar parameters allows
us to perform a complex heterodyne on the data, using the precise
pulsar phase evolution, and down-sample it to $\frac{1}{60}$\,Hz.
Following this heterodyne a Bayesian parameter estimation for the
unknown pulsar parameters $h_0$, $\psi$, $\cos\iota$ and $\phi_0$
can be performed on the massively reduced data set.

The method relies on the accuracy of the signal model, and it is
very important that the phase evolution in the heterodyne is
sufficiently good. Any drift from the true pulsar signal phase could
nullify the search if it becomes too severe. This paper will discuss
how to ensure that this phase model is sufficiently accurate for
pulsars in binary systems and those strongly affected by timing
noise. Most recently, these methods have been used in
Ref.~\cite{Abbott:2006} to obtain limits on the gravitational wave
emission from 78 pulsars using data from the LIGO and GEO\,600 S3
and S4 runs.

\section{Pulsars in binary systems}\label{Binaries}
Of the 1627 pulsars in the ATNF catalogue, 124 are in binary
systems.  The first of these to be discovered was the highly
relativistic pulsar PSR\,J1915+1606 found by Hulse and Taylor in
1974 \cite{HulseTaylor:1975}. Of these 124 pulsars, 98 have spin
frequencies greater than 25\,Hz (out of a total 163 isolated {\it
and} binary pulsars with frequencies greater than 25\,Hz), and 
therefore gravitational wave frequencies $> 50$\,Hz, putting them 
into the sensitive band of the LIGO detectors. Indeed, for the reason 
stated in \S\ref{intro}, the majority of millisecond pulsars are in binary
systems. 

\subsection{Pulsar timing}
A brief discussion on how pulsar timing information is obtained is
relevant here. The majority of pulsars have been both discovered and
monitored in radio. Pulsar surveys, discussed in more detail in
Ref.~\cite{PulsarAstronomy}, use Fourier transform methods to look
for periodic signals in the radio data, taking into account the
effect of interstellar dispersion across the receiver band. Once the
pulsar period has been determined, the radio time series data
can be folded with this cadence to build up the signal-to-noise
ratio of a mean pulse. Once a stable pulse is
obtained\footnote{Individual pulses can vary in shape, but the
summation of many gives a generally stable pulse shape.} the time of
arrival (TOA) can be measured at the peak of the pulse. These pulse
times can then be used to extract more precise information about the
pulsar parameters, including its position and frequency parameters.

The most prevalent tool used to fit timing measurements is the
\tempo software suite \cite{TEMPO}, however others  have been
developed, including \textsc{psrtime} \cite{PSRtime} at Jodrell Bank
Observatory. \tempo requires precise solar system ephemerides,
containing the positions and velocities of the major solar system
bodies, to convert TOAs at a detector to the rest frame of the
pulsar. It computes the pulsar phase at each TOA, $\phi(T_i)$, over
the range of pulsar parameters ($\alpha$, $\delta$, $\nu$,
$\dot{\nu}$, etc), and uses a $\chi^2$ goodness-of-fit statistic to
determine the best model via minimization. A starting point for the
fit is obtained through a rough knowledge of the position and
frequency from the initial discovery, but it can still be quite
complex as there can be many other parameters that could be
contributing. Below it is seen how a pulsar in a binary system
requires a complex model with many more parameters than an isolated
object.

\subsection{Binary pulsar timing}
The majority of pulsars within the ground-based \gw frequency band
are in binary systems, so we must consider the effects of binary
timing carefully.  Techniques to create filters to match a binary
signal (in the frequency domain) have been described before, for
example in Ref.~\cite{DhurandharVecchio:2001},  and these have been
applied to the search for \gws from the neutron star in Sco X-1
\cite{FstatPaper:2006}, but there has been no equivalent treatment
in the time domain, suitable for the search outlined above.
Eq.~(\ref{TimeDelay}) shows the timing corrections needed to take
account of Doppler and relativistic delays of a signal and transform
it to the SSB. Any constant Doppler delays from the pulsar's actual
motion relative to the SSB are unimportant, and the SSB frame can be
considered as the rest frame of the pulsar. For a pulsar in a binary
system however, its motion within the system will need to be taken
into account with a transform from the binary system barycentre to
the pulsar proper time.

The basic transformation and binary models below are summarised in
Ref.~\cite{TaylorWeisberg:1989} and used in the pulsar timing
program \tempo \cite{TEMPO}. The transformation from SSB time $t_b$
to pulsar proper time $T$ follows the form of Eq.~(\ref{TimeDelay})
and is
\begin{equation}\label{SSBtoPPT}
t_b - t_0 = T + \Delta_{\rm{R}} + \Delta_{\rm{E}} + \Delta_{\rm{S}} + \Delta_{\rm{A}},
\end{equation}
where $\Delta_{\rm{R}}$ is the Roemer time delay giving the
propagation time across the binary orbit; $\Delta_{\rm{E}}$ is the
Einstein delay and gives gravitational redshift and time dilation
corrections; $\Delta_{\rm{S}}$ is the Shapiro delay which gives
the gravitational propagation delay due to the signal propagation 
through the curved space-time of the companion; and $\Delta_{\rm{A}}$ is
the aberration delay caused by the pulsar's rotation.

The majority of binary orbits can be well-modelled as Keplerian.
Keplerian orbits are defined by five parameters, $T_0$ - the
time of periastron (closest approach in the binary orbit); $\omega$
- the longitude of periastron; $P_b$ - the orbital period; $e$ - the
orbital eccentricity (where $e = \sqrt{(1-b^2/a^2)}$ and $a$ and $b$
are the semi-major and semi-minor axis of the orbital ellipse
respectively); and $x \equiv (a\sin{i})/c$ is the projected
semi-major axis, with $i$ being the orbital inclination.  For some
of the most extreme binary systems, with rapidly-spinning pulsars in
tight orbits, the maximum (gravitational wave) orbital Doppler
frequency shift can be up to $\sim 0.1$\,Hz.  For example,
PSR\,J0737-3039A with $x=1.415$\,light sec, $P_b = 0.10225$\,days
and $\nu_{\rm gw} = 88.11$\,Hz gives $\Delta\nu_{\rm gw} =
0.089$\,Hz.

We will now consider the three most commonly used models used to
characterise the TOA of pulses from pulsars in binary systems.

\subsubsection{Blandford-Teukolsky model}
The Blandford and Teukolsky model \cite{BlandfordTeukolsky:1976}
(BT) makes no assumptions about the correct theory of gravity.
Instead, it assumes a simple Keplerian orbit with slow precession,
into which additional relativistic effects have been added. Other
phenomena can be taken into account through time derivatives of the
four main orbital elements, excluding $T_0$. 
The BT model has been used to fit data for 53 of the binary pulsars 
with $\nu > 25$\,Hz, and is
the most common model used. Exceptionally, one of these systems is
modelled using the \tempo model BT2P which accommodates three
orbits, the first of which can be relativistic, but the second and
third are Keplerian. The system is a multiple system, described in
Ref.~\cite{Wolszczan:2000}, in which three, or possibly four,
planets orbit the pulsar. Although these additional orbits
complicate the above equations, the standard BT model is sufficient
for our purposes.

\subsubsection{Low eccentricity model}
The second most common model used in fitting radio observations of
binaries is the low eccentricity model (called ELL1 in
T\textsc{empo}) developed in Ref.~\cite{Lange:2001}. It is used as a
fit for pulsars in very low eccentricity orbits ($e\simeq0$) and is
the appropriate model for 38 of our target pulsars with $\nu >
25$\,Hz. An almost circular orbit makes $T_0$ and $\omega$ nearly
degenerate, so these parameters, along with $e$, are replaced with
the non-covariant parameters of the time of the ascending node of
the orbit ($T_{\rm{asc}} \equiv T_0 - \omega{}P_b/2\pi$) and the
first and second Laplace-Lagrange parameters $\eta \equiv
e\sin{\omega}$ and $\kappa \equiv e\cos{\omega}$. The time delays
for this model are defined in \cite{Lange:2001} and T\textsc{empo}.

\subsubsection{Damour-Deruelle model}
The third most common model is that of Damour and Deruelle (DD)
\cite{DamourDeruelle:1986}. This model uses a method for solving the
relativistic two-body problem to post-Newtonian order and is valid
under very general assumptions about the nature of gravity in strong
field regimes. It is useful for highly relativistic systems,
although in only mildly relativistic systems this model should be no
different to the BT model. There are seven pulsars with $\nu
>25$\,Hz in the ATNF catalogue that use this model. This model is again
summarised in \cite{TaylorWeisberg:1989}.

\subsection{Comparison with TEMPO}\label{sec:comparison}
The above three models are all implemented in the pulsar timing
software package T\textsc{empo}. In our search for \gws from binary
systems we also require these additional time corrections to
correctly calculate the phase of the pulsar for heterodyning. Code
to calculate the binary time delays for each model has been adapted
from the \tempo counterparts and is available under CVS in the LIGO
Algorithm Library (LAL) repository \cite{LAL}. Some consistency
tests have been performed between the two codes, which are described
below. It is acknowledged that \tempo has some uncertainties in its
timing models due to simplifying assumptions, and these limit its
accuracy to $\sim$100\,ns. These effects are discussed in
Refs.~\cite{Hobbs:2006a, Edwards:2006}, but as errors at this level
have no significant impact on our search we will neglect them.

\subsubsection{PSR\,J1012+5307}
We performed a validation check on the new LAL code by
demodulating radio data from a known pulsar. With no known
gravitational calibrators, this is a crucial step in the development
of any gravitational wave detection code. A set of TOAs for
PSR\,J1012+5307 obtained with the Effelsberg 100\,m radio telescope
in Bonn, Germany, was kindly supplied by Michael Kramer for
this purpose. This pulsar has the second most circular orbit known
and is therefore fitted by the ELL1 model. The data intermittently
spanned just over 5 years from $2^{\rm{nd}}$ January 2000
to $12^{\rm{th}}$ February 2005 and comprised TOAs (in 
Modified Julian Date format, where MJD = Julian Date$-$2400000.5),
together with the T\textsc{empo}-derived pulsar parameters
(Table~\ref{PSRJ1012+5307}).
\begingroup
\squeezetable
\begin{table}[!htbp]
\caption[The parameters of PSR J1012+5307.]{\label{PSRJ1012+5307} The parameters of PSR J1012+5307.
Values are quoted with $1\sigma$ errors on the final digit in brackets.}
\begin{tabular}{ r | l }
\hline \hline
\multicolumn{2}{ c }{PSR\,J1012+5307} \\
\hline
$\alpha$ & $10^{\rm{h}}12^{\rm{m}}33^{\rm{s}}.43368(1)$ \\
$\delta$ & $53^{\circ}07'02''.5880(2)$ \\
PMRA   & 2.38(3)\,mas/yr \\
PMDEC  & $-25.35$(5)\,mas/yr \\
$\nu$  & 190.267837621884(9)\,Hz \\
$\dot{\nu}$ & $-6.2022(2)\ee{-16}$\,Hz/s \\
$\ddot{\nu}$ & $2.0(3)\ee{-27}$\,Hz/$\rm{s}^2$ \\
Frequency epoch & MJD\,50700 \\
Dispersion measure & 9.0233(7)\,${\rm{cm}}^{-3}$\,pc \\
Observing Frequency  & 1408.6\,MHz \\
Binary model & ELL1 \\
$x$ & 0.581817(1)\,s \\
$P_b$ & 0.6046727136(2)\,days \\
$T_{\rm{asc}}$ & MJD\,50700.0816289(4) \\
$\eta$  &  $7(4)\ee{-7}$ \\
$\kappa$  &  $-1(40)\ee{-8}$ \\
\hline \hline
\end{tabular}
\end{table}
\endgroup
Some minor transformations are necessary to convert TOAs measured at
the telescope to the GPS time stamps used in gravitational wave data
analysis software. First the raw TOAs are corrected for the drifts
between the hydrogen maser clock at Effelsberg and coordinated
Universal Time of the National Institute of Science and Technology
UTC(NIST) reference. This correction (supplied with the data) was
typically a few microseconds. The difference between UTC(NIST) and
UTC has been less than $\pm 100$\,ns since $6^{\rm{th}}$ July,
1994 \cite{BIPM} and was neglected for this work. The conversion 
between the time scales therefore becomes
\begin{equation}
t_{\rm{GPS}} = (t_{\rm{UTC(MJD)}} -
44244\,{\rm{days}})\times86400\,{\rm{s}} + L,
\end{equation}
where the 44244 corresponds to the MJD of the GPS time epoch
($1^{\rm{st}}$ January, 1980) and $L$ is the accumulated number of
leap seconds included in the definition of UTC.  For the time-span
of these TOAs,  $L=13$.

The TOAs can now be corrected for interstellar dispersion time delay \cite{PulsarAstronomy}
\begin{equation}
\Delta{}t_{\rm{disp}} =
4.149\ee{3}\,{\rm{MHz}}^2\,{\rm{pc}}^{-1}\,{\rm{cm}}^3\,{\rm
s}\times{\rm{DM}}/f^2\,{\rm s},
\end{equation}
where DM is the dispersion measure in ${\rm{cm}}^{-3}$\,pc and $f$
is the radio observation frequency in MHz (see
Table~\ref{PSRJ1012+5307} for values). This correction is subtracted
from the TOAs to give observations at infinite frequency with no
dispersion.

One of the major differences between our binary time domain code and
the \tempo code is the time system used. All epochs in \tempo are
defined as MJD Barycentric Dynamical Time (TDB - a timescale
generally used for ephemerides referenced to the solar solar
barycentre) whereas the general reference time for our \gw data is
GPS time. Epochs therefore have to be converted to GPS time on the
TDB timescale. This TDB timescale is related to Terrestrial Time (TT
- formerly Terrestrial Dynamical Time TDT), which represents a time
consistent with relativity for an observer on the Earth's surface,
by a small factor, ${\rm TDB} = {\rm TT} + \delta{}t$, no greater
than a couple of milliseconds and given by
\begin{equation}
\delta{}t = 0.001\,658\,{\rm s}\times\sin{\Phi} + 0.000\,014\,{\rm s}\times\sin{2\Phi},
\end{equation}
where $\Phi = 357.53^{\circ} + 0.985\,600\,28^{\circ}({\rm MJD} -
51\,544.5)$ is the mean anomaly, or phase, of the Earth's orbit at
the given Modified Julian Date \cite{Times}. TT is offset from
International Atomic Time (TAI), so that TT = TAI + 32.184
seconds\footnote{There are many definitions of time used in
astronomy and very careful attention of which one is being used and
how to convert between them is essential when high precision timings
are being made. A good guide for these is \cite{Times}.}.
The conversion thus goes $t_{\rm TDB(GPS)} = (t_{\rm
TDB(MJD)}-44244) - 51.184 - \delta{}t$, where the 51.184\,s
comprises  the 32.184\,s difference between TT and TAI and 19
second difference between TAI and GPS.

The LAL code to calculate the SSB time delay uses the pulsar's
position, the telescope position and a solar system
ephemeris \cite{JPLEphemeris} and was used for
each pulsar TOA to correct to the time at the SSB. This code was
written by Curt Cutler and has been independently tested against
\tempo \cite{Abbott:2004, Dupuis:2004} showing no more than
$4\,\mu$s difference between the two. For a pulsar with a gravitational 
wave frequency at 1\,kHz a 4\,$\mu$s timing error would give a 
phase offset between any signal and our model template of 0.025\,rads. 
This would reduce our sensitivity to a signal by of order $1-\cos{0.025} 
\approx 3\ee{-4}$, which is negligible.

\begin{figure}[!htbp]
\begin{center}
\includegraphics[width=0.45\textwidth]{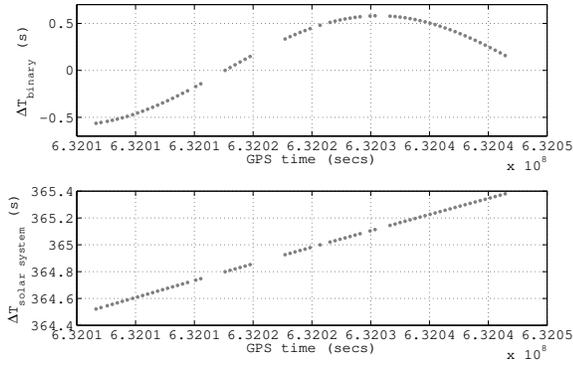}
\caption[The binary and solar system time delays for PSR\,J1012+5307.]{The binary and solar system
time delays calculated for PSR\,J1012+5307 over a part of a binary orbit.}\label{J1012+5307DT}
\end{center}
\end{figure}

Once corrected to the SSB the TOAs  are further corrected to the
pulsar proper time by calculating the time delays in the binary
system using the binary system parameters (see
Table~\ref{PSRJ1012+5307}). The binary and solar system time delays
for a selection of TOAs covering part of the binary orbit are shown
in Fig.~\ref{J1012+5307DT}.

\begin{figure}[!htbp]
\begin{center}
\includegraphics[width=0.45\textwidth]{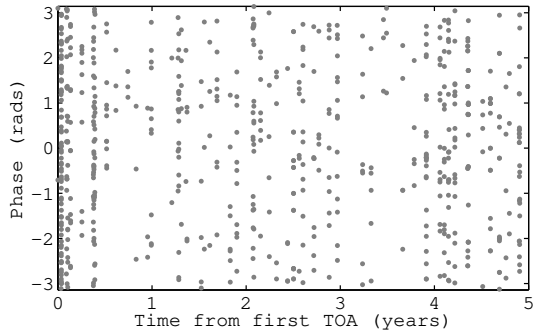}
\caption{The modulus of the pulsar phase at each TOA over a 5 year period
with no binary time delays included.}\label{NoBinaryDelays}
\end{center}
\end{figure}

\begin{figure}[!htbp]
\begin{center}
\includegraphics[width=0.45\textwidth]{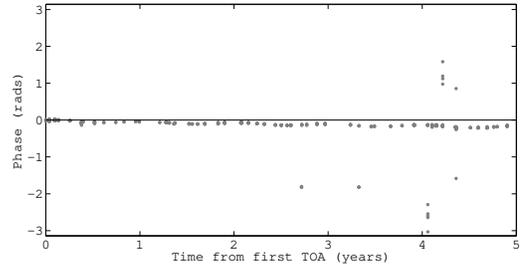}
\caption{The modulus of the pulsar phase at each TOA over a 5 year
period corrected for the binary delay.}\label{J1012+5307Phase}
\end{center}
\end{figure}

Once these corrections to the TOAs are applied we can compare the
LAL barycentring codes with \tempo by inserting the {\sc Tempo}-derived
pulsar parameters into the LAL barycentring routines and examining
the predicted phase at the corrected TOAs.  TOAs converted
incorrectly by the barycentring codes would show up as a phase
drift. The phase at each TOA was calculated using the supplied
frequency and frequency derivatives in Eq.~(\ref{PhaseTaylorExp}),
with $\phi_0 = 0$ and the frequency epoch as $t_0$.
Fig.~\ref{NoBinaryDelays} demonstrates the serious effect of
neglecting these binary time delays.
In contrast, Fig.~\ref{J1012+5307Phase} shows how the TOAs
barycentred using our code stay well in phase over the observation
time when the binary delays are included, with a residual phase
slope of $\sim 0.04$\,rads/yr.
A yearly periodicity is also present possibly showing up the slight
difference in the LAL solar system barycentring code and
T\textsc{empo}, although these effects are at a very low level.
Several points clearly show large phase residuals and correspond to
times when the level of noise on the TOA measurements was high.

\begin{figure}[!htbp]
\begin{center}
\includegraphics[width=0.45\textwidth]{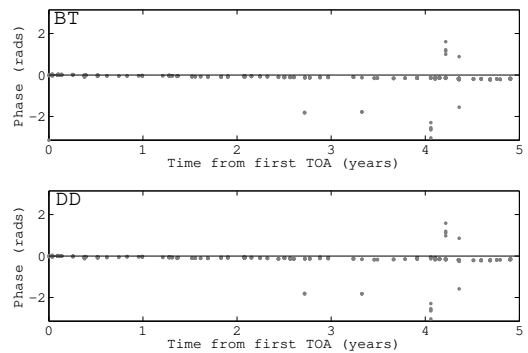}
\caption{The modulus of the pulsar phase at each TOA over a 5 year
period for the BT and DD models, corrected for the binary
delay.}\label{J1012+5307BT_DDPhase}
\end{center}
\end{figure}

The parameters for PSR\,J1012+5307 were generated using the ELL1
model, so the above test only checked the ELL1 code. We can also
check the two other models by converting $T_{\rm{asc}}$ to $T_0$ and
the Laplace-Lagrange parameters $\kappa$ and $\eta$ to $e =
\sqrt{(\kappa^2+\eta^2)}$ and $\omega$. This pulsar has a low
eccentricity, so $T_0$ can be set equal to $T_{\rm{asc}}$ and $e$
and $\omega$ set to zero for practical purposes. Doing this we can
again produce the phase plots for the BT and DD models
(Fig.~\ref{J1012+5307BT_DDPhase}).
The phase is again well described for these two models, with the
slope and periodicity still present. This suggests that the slope
and periodicities are not caused by the binary timing correction
code (as each model is independent), but may be a results of slight
errors in the other timing corrections, the solar system
barycentring code, the ephemerides, or the parameters used.

\begin{figure}[!htbp]
\begin{center}
\includegraphics[width=0.45\textwidth]{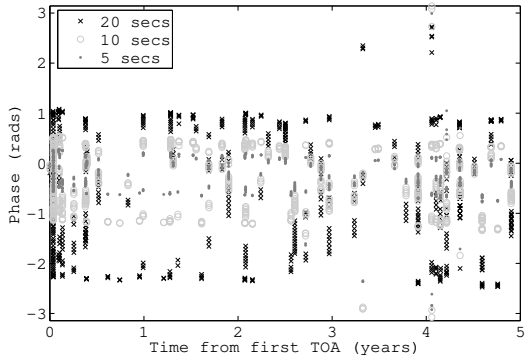}
\caption{The modulus of the pulsar phase at each TOA over a 5 year period with $T_{\rm asc}$ offset
from its true value.}\label{TascOffset}
\end{center}
\end{figure}

The effects of inaccuracies in these binary parameters can be shown
by offsetting one from its true value. In Fig.~\ref{TascOffset} the
true value of the $T_{\rm asc}$ parameter has been offset by 5, 10
and 20 seconds and the phase at each TOA recalculated.
Even for a 5\,s mismatch we start to depart from the true phase by
up to 0.5 radians. This would not be disastrous for the analysis but
would degrade the search. For a 20\,s offset we start to see phase
errors of up to 2 radians.  At this level, all sensitivity is lost.
In real analyses as in Ref.~\cite{Abbott:2006} the effect of any
errors on the parameters, as to whether they could cause our
heterodyne phase to depart significantly from the signal phase, is
thoroughly checked.

\subsubsection{Direct check against TEMPO}
\begin{figure*}[!htbp]
\begin{center}
\includegraphics[width=0.95\textwidth]{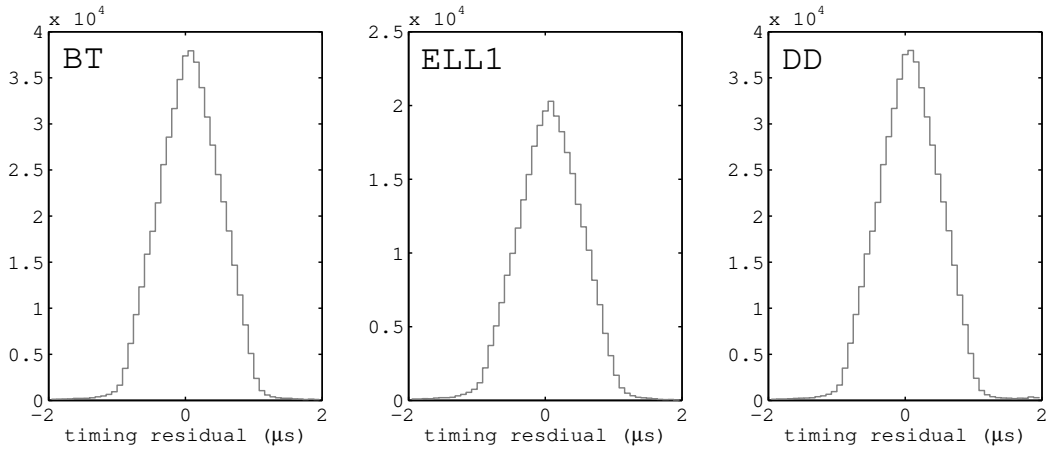}
\caption[Timing residuals between our code and
T\textsc{empo}.]{Timing residuals between the pulsar phase as
predicted by \tempo and that computed with our binary code for 100
random pulsars for each binary model.}\label{TEMPOComparison}
\end{center}
\end{figure*}

In common with the solar system barycentring code \cite{Dupuis:2004,
Abbott:2005}, the binary timing code was tested directly against
T\textsc{empo}. \tempo can be run in \emph{predictive mode}, to use
a set of pulsar parameters to predict the pulsar phase over a period
of time. This predicted phase can be then be compared with that
calculated using our binary timing code. This was done for each
model with a set of 100 randomly generated binary pulsar systems
over a period of 100 days. The detector location was set to be at
the SSB, so the solar system time delay errors would not be
included. Histograms of the time residuals between the codes are
shown in Fig.~\ref{TEMPOComparison} for each model.
These show that the time difference between the two codes is
generally less than $\pm1\,\mu$s, which is sufficiently small to
ensure any signal and template remain in phase over the frequency
range considered.

\section{The problem of timing noise}\label{TimingNoise}
A rather different issue that if unaddressed could potentially cause 
problems for our analysis is that of timing noise. 

Pulsars are generally very stable over periods of several days, but
there are phenomena which can cause deviations in this timing
stability. With the very high accuracy of pulsar timing any random
timing irregularities will start to become evident. These phenomena
show up as glitches and timing noise. Timing noise has been known
about since the early days of pulsar observations and represents a
random walk in phase, frequency or frequency derivative of the
pulsar about the regular spin-down model given in Eq.~(\ref{PhaseTaylorExp})
\cite{CordesHelfand:1980}. If our search method cannot take account of these 
phase deviations they could reduce our sensitivity or even invalidate 
the search. It is therefore important to see firstly if timing noise 
produces large enough phase deviations to cause a problem for our method 
and if so how it can be countered.

The strength of this effect has been quantitatively defined in
Ref.~\cite{CordesHelfand:1980} as the activity parameter $A$, as
referenced to that of the Crab pulsar, and in
Ref.~\cite{Arzoumanian:1994} as the stability parameter $\Delta_8$.
There is however no real consensus on how to quantitatively define a 
measure of the level of timing noise, with
the magnitude and sign of $\ddot{P}$ maybe providing the measure
which includes the least other assumptions. A thorough study of
timing noise, comparing and contrasting the various measures used,
is given in Ref.~\cite{Hobbs:2006} (also see Refs.~\cite{Hobbs:2004,
Hobbs:2005}). There is a definite correlation between these
parameters ($A$ and $\Delta_8$) and the pulsar's spin-down rate,
therefore possibly the pulsar's age. Young pulsars, like the Crab
pulsar, generally show the most timing noise activity.

The Crab pulsar is the youngest known pulsar targeted by current \gw
detectors and it is important to considered how timing noise may be countered
for this pulsar.  A method was first proposed in
Ref.~\cite{PitkinWoan:2004} and has been used in the analysis of
Abbott {\it et al.} \cite{Abbott:2005}, but has not previously been
described in detail. A brief look into the effects of timing noise
for other pulsars will be discussed later.

\subsection{Timing noise in the Crab pulsar}
The pulsar (J0534+2200) in the Crab nebula (M1) has undergone intense 
study since its discovery in 1968. Its parameters are given in Table~\ref{CrabParams}.
\begingroup
\squeezetable
\begin{table}[!htbp]
\caption{\label{CrabParams} The parameters of the Crab pulsar as 
calculated from the Jodrell Bank monthly ephemeris 
\cite{CrabEphemeris}.}
\begin{center}
\begin{tabular}{ c | l }
\hline \hline
\multicolumn{2}{ c }{PSR\,J0534+2200} \\
\hline
Right ascension $\alpha$ & $05^{\rm{h}} 34^{\rm{m}} 31^{\rm{s}}.973$ \\
Declination $\delta$ & $22^{\circ} 00' 52''.06$ \\
proper motion in $\alpha$ & $-13$ mas/yr \\
proper motion in $\delta$ & 7 mas/yr \\
Position epoch & MJD\,40675 \\
$\nu$ & 29.7670971390\,Hz \\
$\dot{\nu}$ & $-3.72633\ee{-10}$\,Hz/s \\
$\ddot{\nu}$ & $1.1429\ee{-20}$\,Hz/${\rm{s}}^2$ \\
Frequency epoch & MJD\,53993 \\
Distance & 2.0 kpc \\
\hline \hline
\end{tabular}
\end{center}
\end{table}
\endgroup
It has an observed age of 951 years (the
formation of the Crab nebula is associated with a supernova observed
in AD 1054) and a spin-down age of $-\nu/2\dot{\nu} = P/2\dot{P} =
1250$\,years. Analyses of long-term timing observation of the Crab 
pulsar are given in Refs.~\cite{Lyne:1993, Wong:2001}. These analyses 
show some of the timing features which make the Crab pulsar such an 
interesting object: the timing noise and glitches.

\begin{figure}[!htbp]
\begin{center}
\includegraphics[width=0.45\textwidth]{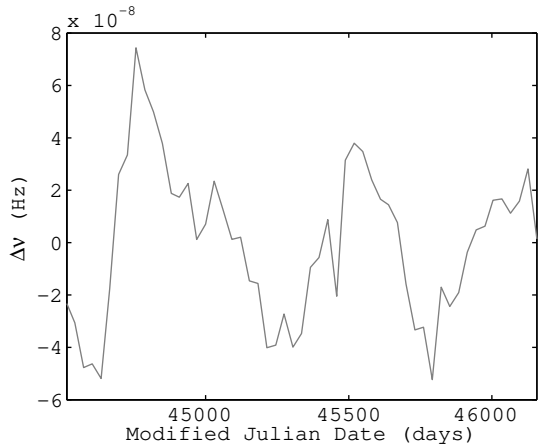}
\caption[Crab pulsar frequency timing noise.]{The timing noise in the frequency of the Crab pulsar
after removing a quadratic fit to the frequency as given in the Jodrell Bank
ephemeris.
}\label{CrabTimingNoise}
\end{center}
\end{figure}

Since 1982 there has been a regular monitoring program of the Crab
pulsar at Jodrell Bank Observatory, and timing ephemerides from this
are publicly available online \cite{CrabEphemeris}. The ephemeris
gives the pulsar frequency and frequency derivative and associated
errors, and the associated epoch. The epochs, generally given on the
$15^{\rm{th}}$ of each month, represent the time of the peak of the
first pulse after midnight on that day. They therefore represent
zero of modulus phase of the electromagnetic pulse. Notes are given
in the event of a timing irregularity or glitch being observed. The
Crab's timing noise can be clearly seen in the ephemeris once a
best-fit quadratic timing model (Eq.~(\ref{PhaseTaylorExp})) has been
subtracted (see Fig.~\ref{CrabTimingNoise}). The section of data
used was chosen to be free of glitches as these can be much larger
than any timing noise frequency deviations. 
Fig.~\ref{CrabTimingNoise} compares well with that given in
Ref.~\cite{Lyne:1993}, although some difference can be expected due
to the different lengths of data and epochs used in the fitting. It
can be seen that on scales of several months there is quite a large
variation in the timing residual. It is
shown in Ref.~\cite{Lyne:1993} that on smaller time scales the variation
is far smoother. It is important to consider  whether timing noise
at this level will cause a loss in sensitivity in any search for
gravitational wave emission from the Crab pulsar. Jones \cite{Jones:2004}
constructs a decoherence timescale, $T_{\rm{decoherence}}$, defined
as the time over which the timing noise will cause the phase to
deviate by 1 radian from the second order Taylor expansion of phase.
This makes use of the ``activity parameter'' and is calculated  to
be $\sim 2.6$\,yr for the Crab pulsar. Though useful as an
indicator, this statistic does not take account of permanent changes
to spin-down caused by glitches or other secular variations in the
pulsar.

Although it is reasonable that a third order fit to the entire data
would improve the model, there is no need.  The Crab pulsar
ephemeris provides timing every month, which is sufficient to track
the phase excursions in the timing noise. All that is necessary is
that these be interpolated between the ephemeris times. By using the
phase, frequency and frequency derivative for each entry in the
ephemeris as boundary conditions to a set of simultaneous equations
the full phase evolution between each month can be calculated,
giving a fifth order polynomial,
\begin{eqnarray}\label{5thOrderPhase}
\phi_{5^{\rm th}}(T) = \phi_0 + 2\pi\bigg\{\nu_0(T-t_0) + \frac{1}{2}\dot{\nu_0}(T-t_0)^2+
\nonumber \\
\frac{1}{6}\ddot{\nu_0}(T-t_0)^3 + \frac{1}{24}\dddot{\nu_0}(T-t_0)^4 +
\frac{1}{120}\ddddot{\nu_0}(T-t_0)^5\bigg\}.
\end{eqnarray}
Indeed, for much of the time even this method is unnecessarily
complicated and a simple linear interpolation between months would
be sufficient.

These corrections can be included in the method detailed in
\S\ref{sec:method} as an extra heterodyne step
\cite{PitkinWoan:2004}. The initial heterodyne (described in
\S\ref{sec:method}), uses a third order fit to the the phase with
values of $\nu$ and $\dot{\nu}$ taken from the ephemeris at the
closest time \emph{before} the timestamp on the data to be analysed,
and $\ddot{\nu}$ taken from the ATNF pulsar catalogue value given in
Table~\ref{CrabS2Params}. Then, assuming that any \gw signal would
show the same timing noise (see Refs.~\cite{Jones:2004} and
\cite{Abbott:2006} for discussions of this), we apply a second
heterodyne to the data $B_k$ using the phase difference between
Eqs.~(\ref{PhaseTaylorExp}) and (\ref{5thOrderPhase})
\begin{equation}\label{extraheterodyne}
B_{k}' = B_ke^{-i2[\phi_{5^{\rm th}}(T) - \phi(T)]},
\end{equation}
where the factor of two in the phase is due to the \gw frequency
being twice the spin frequency. This step can be performed on the
data after down-sampling as the rate of change of this phase
difference will be very low.

\begin{figure}[!htbp]
\begin{center}
\includegraphics[width=0.45\textwidth]{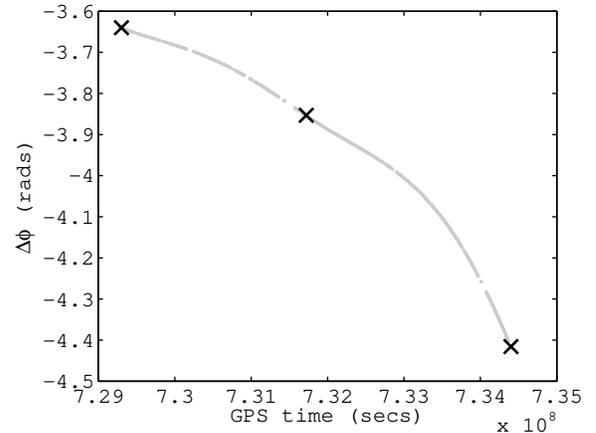}
\caption[Crab timing noise phase check during S2.]{The grey points show the phase difference between
that used in the initial heterodyne and that interpolated from a fifth order fit to the ephemeris.
The black crosses show just the phase difference between the initial heterodyne and the individual
ephemeris values.}\label{S2TimingNoisePhaseDiff}
\end{center}
\end{figure}

The effect of this extra heterodyne can be seen in search for a
signal from the Crab in the S2 data. This science run of the LIGO
interferometers lasted approximately two months and overlapped three
entries in the Crab pulsar ephemeris. The S2 run started on
$14^{\rm{th}}$ Feb 2003, so values of the frequency and spin-down
used in the initial heterodyning were chosen to be those given in
the first ephemeris entry prior to the run (15th Jan 2003). The
second derivative was set to be that taken from the ATNF catalogue.
The values are shown in Table~\ref{CrabS2Params}.
\begingroup
\squeezetable
\begin{table}[!htbp]
\caption{\label{CrabS2Params} The parameters used in the initial heterodyne stage of
the Crab pulsar analysis for S2.}
\begin{center}
\begin{tabular}{ c | l }
\hline \hline
\multicolumn{2}{ c }{PSR\,J0534+2200} \\
\hline
$\nu$ & 29.8102713888\,Hz \\
$\dot{\nu}$ & $-3.736982\ee{-10}$\,Hz/s \\
$\ddot{\nu}$ & $1.2426\ee{-20}$\,Hz/$\rm{s}^2$ \\
Frequency epoch & GPS\,726624013 \\
\hline \hline
\end{tabular}
\end{center}
\end{table}
\endgroup
Once the data were produced the ephemeris values were used to
calculate the phase given in Eq.~(\ref{5thOrderPhase}). The
difference between the initial heterodyne phase and the $5^{\rm th}$
order phase is shown in Fig.~\ref{S2TimingNoisePhaseDiff}.
This phase difference is used in the extra heterodyne to remove the
variation. It can be seen in Fig.~\ref{S2TimingNoisePhaseDiff} how a
linear fit between ephemeris values would be acceptable for these
times, with only small deviations in phase from the fifth order fit.
The black crosses in Fig.~\ref{S2TimingNoisePhaseDiff} provide the
first step in checking the code used for the extra heterodyne stage.
The red points represent the phase difference used in our extra
heterodyne step (Eq.~(\ref{extraheterodyne})) to heterodyne each S2
data point as calculated using our code, whereas the black crosses
just show the phase difference between the initial heterodyne and
the individual Crab pulsar ephemeris data points. The fact that
these overlap provides a check that the heterodyne code is producing
the correct phase difference.

\begin{figure}[!htbp]
\begin{center}
\includegraphics[width=0.45\textwidth]{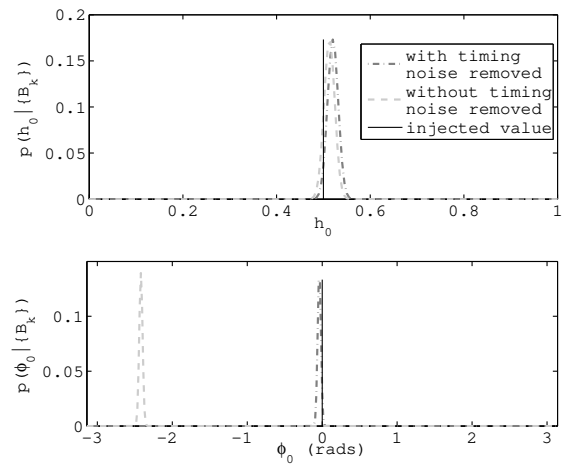}
\caption{The extracted pdfs for $h_0$ and $\phi_0$ for a simulated signal from the
Crab pulsar over the period of S2 with and without timing noise removed.}\label{S2CrabInjection}
\end{center}
\end{figure}

If we simulate a signal from the Crab pulsar over the period of S2,
with parameters $h_0 = 0.5$, $\phi_0 = 0.0$, $\psi = 0.0$ and
$\iota=\pi$, we can see how including a timing noise heterodyne step
affects the parameter estimation. Fig.~\ref{S2CrabInjection} shows
the extracted probability distribution functions (pdfs) of $h_0$ and
$\phi_0$ for the signal with and without the timing noise removed.
There is very little difference between the amplitudes for the two
cases because the slope of the phase difference $\Delta\phi$ are not
too steep over the period of S2.
However, the extracted value of the phase is strongly affected,
mainly due to the the phase offset between the start of S2 and the
epoch of the initial heterodyne parameters seen in
Fig.~\ref{S2TimingNoisePhaseDiff}.

\begin{figure}[!htbp]
\begin{center}
\includegraphics[width=0.45\textwidth]{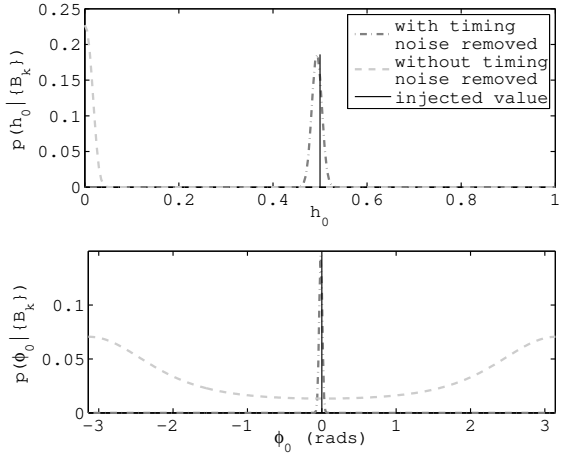}
\caption{The extracted pdfs for $h_0$ and $\phi_0$ for a simulated signal from the
Crab pulsar over the period of S3 with and without performing the extra
heterodyne.}\label{S3CrabInjection}
\end{center}
\end{figure}

\begin{figure}[!htbp]
\begin{center}
\includegraphics[width=0.45\textwidth]{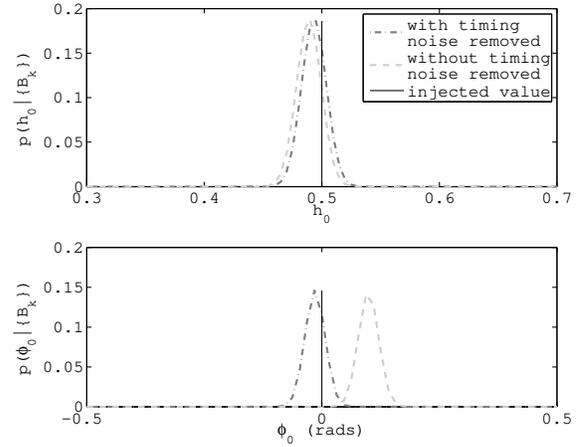}
\caption[Extracted pdfs of simulated Crab signal during S3.]{The
extracted pdfs for $h_0$ and $\phi_0$ for a simulated signal from
the Crab pulsar over the period of S3 with and without timing noise
removed for initial heterodyne values obtained from a fit to the
ephemeris over 2003 (see
Table~\ref{CrabParams2003}).}\label{S3CrabInjection2}
\end{center}
\end{figure}

We can simulate a Crab pulsar signal and analyse it with and without
the timing noise heterodyne step over longer periods than just S2 to
show its importance. The same process as above has been carried out
over the period of the S3 run, using the same initial heterodyne
parameters and pulsar injection parameters. The extracted pdfs are
shown in Fig.~\ref{S3CrabInjection} and demonstrate that without the
extra heterodyne the signal is completely lost.
The fact that this signal is not seen without the extra heterodyne
may appear at odds with the approximate 2.6 year decoherence time
stated above, as S3 was only about 8 months after S2. However, the
initial heterodyne values used were still those from the Crab
ephemeris closest to the start of S2 and not those from a more
general fit to the data over an extended period, as was used to
calculate $T_{\rm{decoherence}}$, so it is not the timing noise
causing the decoherence in this case, but badly chosen initial
heterodyne parameters. This does however demonstrate the importance
of having well-defined heterodyne parameter for all pulsars.  If we
perform a fit to the Crab pulsar ephemeris over the whole of 2003
(see Table~\ref{CrabParams2003} for fit values), when the S2 and S3
runs took place, we can again check the impact of the timing noise
in S3 (see Fig.~\ref{S3CrabInjection2}).
\begingroup
\squeezetable
\begin{table}[!htbp]
\caption{\label{CrabParams2003} The parameters of the Crab pulsar
for a fit to second order in frequency over the period of 2003 using
monthly ephemeris data.}
\begin{center}
\begin{tabular}{ c | l }
\hline \hline
\multicolumn{2}{ c }{PSR\,J0534+2200} \\
\hline
$\nu$ & 29.81027139567395\,Hz \\
$\dot{\nu}$ & $-3.73698\ee{-10}$\,Hz/s \\
$\ddot{\nu}$ & $1.07086\ee{-20}$\,Hz/$\rm{s}^2$ \\
Frequency epoch & GPS\,726624013.0597030 \\
\hline \hline
\end{tabular}
\end{center}
\end{table}
\endgroup
We see that, with an extended fit for the Crab parameters, timing
noise makes little difference over S3, although a slight phase
offset is present.

\subsection{Timing noise in other pulsars}
For the majority of pulsars timing noise is most prominent in the
second derivative of frequency, but for millisecond pulsars this
value is often so small as to be unmeasurable. For these pulsars,
the value of the  $\Delta_8$ parameter can be used to estimate the
cumulative phase contribution of timing noise via the empirical
relationship between $\dot{P}$ and $\Delta_8$ given in
Ref.~\cite{Arzoumanian:1994}. Abbott et al.\ \cite{Abbott:2006} used
this as a test of the coherence of the pulsars phases over the
periods of the LIGO and GEO\,600 S3 and S4 data runs. The generally
small values of $\ddot{\nu}$ for the majority of other pulsars
suggest that there is not significant timing noise.

There is however, one {\it known} interesting target pulsar within
our frequency range that has similar properties to the Crab pulsar.
This pulsar, PSR\,J0537-6910, is young and has the second largest
spin-down after the Crab pulsar, making it a very interesting
candidate for our search. It is distant, lying in the Large
Magellanic Cloud, and has so far only been observed in X-rays. It is
also a prolific glitcher, showing high levels of timing noise
\cite{Marshall:2004}. As dedicated time on the {\it Rossi X-ray
Timing Explorer} satellite is required to time this pulsar, it has
not been observed as regularly as the Crab pulsar and a comprehensive
ephemeris does not exist. A regularly updated ephemeris would
however allow us to track the inter-glitch phase and perform an
analysis similar to that of the Crab pulsar.

\section{Conclusions}
We have shown how to take account of time delays due to the motion
of pulsars within binary systems in searches for \gws. This is
particularly important as a large number of the pulsars within the
sensitivity band of current interferometric \gw detectors are in
binaries.  However, as discussed in Ref.~\cite{Abbott:2006}, we have
demonstrated the importance of knowing the binary parameters and
their covariances.

At present our search code contains the three main binary models
described above. These are sufficient for the majority of pulsars,
although \tempo contains many more models which could be
incorporated in the future if needed.

For the cases where large levels of timing noise are seen it is
important that up-to-date parameters are obtained allowing us to
track the rotational variations. Fortunately, the Crab pulsar is
constantly monitored in radio and therefore has a very
well-determined  phase evolution. Other pulsars in our band are not
so well monitored and this is especially pertinent if we want to
target young pulsars (generally the best candidates for \gws due to
there large spin-down rates), because these will generally also be
the most affected by timing noise. In particular, the young X-ray
pulsar PSR\,J0537-6910 \cite{Marshall:2004} has high levels of
glitch-induced timing noise, and an ability to regularly monitor
this object would allow a more detailed study of this prime
gravitational wave candidate.

\begin{acknowledgments}
The authors are very grateful to Michael Kramer for supplying pulsar
timing information and useful discussions on its use. We would also
like to acknowledge members of the LIGO Scientific Collaboration for
useful advise during the writing of this manuscript. This work is to
a large extent an extension of the work of R\'ejean Dupuis and to
him we extend many thanks. This work has been supported by the UK
Particle Physics and Astronomy Research Council.
\end{acknowledgments}

\end{document}